\documentclass[prd,aps]{revtex4}
\usepackage{graphicx}
\begin{document}
\title{Critical collapse of a massive vector field}
\author{David Garfinkle \thanks{%
Email: garfinkl@oakland.edu}}
\affiliation{
\centerline{Department of Physics, Oakland University,
Rochester, Michigan 48309}}

\author{Robert Mann \thanks{%
Email: rbmann@sciborg.uwaterloo.ca }}
\affiliation{
\centerline{Guelph-Waterloo Physics Institute, Department of Physics,
University of Waterloo, Waterloo, Ontario, Canada N2L 3G1
}}

\author{Chris Vuille \thanks{%
Email: cvuille@yahoo.com }}
\affiliation {Department of Physical Sciences, Embry-Riddle
Aeronautical University, Daytona Beach, FL 32114
\centerline{
}} \null\vspace{-1.75mm}

\begin{abstract}
We perform numerical simulations of the critical gravitational collapse of a
massive vector field. The result is that there are two critical solutions.
One is equivalent to the Choptuik critical solution for a massless scalar
field. The other is periodic.
\end{abstract}


\maketitle

\section{Introduction}

Critical gravitational collapse was first found by Choptuik\cite{matt} in
simulations of a spherically symmetric massless scalar field. A natural
question to pose is then how critical collapse behaves when the scalar field
has a mass, since this will introduce a characteristic length that
destroys the scale invariance of the field equations.
This question was studied by Brady \textit{et. al.}\cite{pat} The results of
reference\cite{pat} show that there are two critical solutions:
one which is essentially the Choptuik critical solution for the massless
scalar field, and another which is a periodic solution first found
by Seidel and Suen \cite{edandwaimo}. At first it might seem
puzzling that the Choptuik solution can be a critical solution for both the
massless and massive scalar field. The resolution of this conundrum
is that as the singularity is approached in the Choptuik critical solution,
the amplitude of the scalar field remains bounded while its gradient
diverges. In the stress energy tensor, the mass terms are associated
with the amplitude of the field, while other terms are associated
with its gradient.
So as the singularity is approached the
mass terms in the stress energy become negligible.

Given the results of reference\cite{pat} one might conjecture that similar
behavior occurs in the case of a spherically symmetric massive vector field: 
\textit{i.e.} that there is a critical solution for which the mass of the
vector field becomes negligible. However, this conjecture involves a
paradox: a massless vector field is just a Maxwell field, and a
spherically symmetric Maxwell field has no degrees of freedom. Therefore
there is no gravitational collapse (and thus no critical solution) of a
spherically symmetric massless vector field. What then is the critical
behavior of a massive vector field? 

\bigskip 

In this paper we consider this question. We perform numerical
simulations of the collapse of a spherically symmetric massive vector field.
The equations and numerical methods are presented in section 2. Results are
given in section 3 and conclusions in section 4.

\section{Equations and Numerical Methods}

A massive vector field is described by the Proca Lagrangian 
\begin{equation}
\mathcal{L}=-{\frac{1}{4}}{F_{ab}}{F^{ab}}-{\frac{1}{2}}{\mu ^{2}}{A_{a}}{%
A^{a}}  \label{procaL}
\end{equation}%
where ${F_{ab}}={\nabla _{a}}{A_{b}}-{\nabla _{b}}{A_{a}}$ and $\mu $ is a
constant. Note: throughout, we use the conventions of Wald\cite{bob} and in
particular use a metric with signature $(-,+,+,+)$. 
Note that were we to employ the opposite signature, the sign of one of 
the terms in the Proca Lagrangian would have to be changed
(see \textit{e.g.}\cite{iz}). The
equation of motion that follows from equation (\ref{procaL}) is 
\begin{equation}
{\nabla _{a}}{F^{ab}}-{\mu ^{2}}{A^{b}}=0  \label{divF}
\end{equation}%
from which it follows that 
\begin{equation}
{\nabla _{a}}{A^{a}}=0  \label{divA}
\end{equation}%
It also follows from equation (\ref{procaL}) that the Einstein field
equation is 
\begin{equation}
{G_{ab}}=2{F_{ac}}{{F_{b}}^{c}}+2{\mu ^{2}}{A_{a}}{A_{b}}-{g_{ab}}\left( {%
\frac{1}{2}}{F_{cd}}{F^{cd}}+{\mu ^{2}}{A_{c}}{A^{c}}\right) 
\label{Einstein}
\end{equation}

We now specialize to spherical symmetry. We employ two different
methods to simulate the Einstein-Proca system: a Cauchy method using
polar-radial coordinates and a characteristic method using the coordinates
used by Christodoulou\cite{dc} to treat the Einstein-scalar system. The
metric in polar-radial coordinates takes the form 
\begin{equation}
d{s^{2}}=-{\alpha ^{2}}d{t^{2}}+{a^{2}}d{r^{2}}+{r^{2}}(d{\theta ^{2}}+{\sin
^{2}}\theta d{\phi ^{2}})  \label{trmetric}
\end{equation}%
Define the quantities $X$ and $W$ by 
\begin{eqnarray}
X &\equiv &{\frac{a}{\alpha }}{A_{t}}  \label{Xdef} \\
W &\equiv &{\frac{1}{{\alpha a}}}({\partial _{t}}{A_{r}}-{\partial _{r}}{%
A_{t}})  \label{Wdef}
\end{eqnarray}%
From the definition of $W$ we find 
\begin{equation}
{\partial _{t}}{A_{r}}=\alpha aW+{\partial _{r}}\left( {\frac{\alpha }{a}}%
X\right)   \label{evolveAr}
\end{equation}%
Then from equation (\ref{divA}) we have 
\begin{equation}
{\partial _{t}}X={\frac{1}{{r^{2}}}}{\partial _{r}}\left( {\frac{\alpha }{a}}%
{r^{2}}{A_{r}}\right)   \label{evolveX}
\end{equation}%
From equation (\ref{divF}) we find the following constraint equation for $W$ 
\begin{equation}
{\frac{1}{{r^{2}}}}{\partial _{r}}({r^{2}}W)+{\mu ^{2}}X=0  \label{constrW}
\end{equation}

Equations (\ref{evolveAr}), (\ref{evolveX}) and (\ref{constrW}) provide the
evolution equations for the matter field. In spherical symmetry the metric
has no degrees of freedom. Therefore the metric functions $\alpha $ and $a$
are given by ``constraint'' equations once the matter fields are known. To
find the appropriate constraint equations, note that for a metric of the
form given in equation (\ref{trmetric}) the $tt$ and $rr$ components of the
Einstein tensor are 
\begin{eqnarray}
{G_{tt}} &=&{\frac{{\alpha ^{2}}}{{{r^{2}}{a^{3}}}}}\left[ -a+{a^{3}}+2r{%
\frac{{\partial a}}{{\partial r}}}\right]   \label{Gtt} \\
{G_{rr}} &=&{\frac{1}{{{r^{2}}\alpha }}}\left[ (1-{a^{2}})\alpha +2r{\frac{{%
\partial \alpha }}{{\partial r}}}\right]   \label{Grr}
\end{eqnarray}%
It then follows from equations (\ref{Gtt}) and (\ref{Einstein}) that 
\begin{equation}
{\frac{1}{a}}{\frac{{\partial a}}{{\partial r}}}+{\frac{{{a^{2}}-1}}{{2r}}}={%
\frac{r}{2}}\left[ {a^{2}}{W^{2}}+{\mu ^{2}}({X^{2}}+{A_{r}^{2}})\right] 
\label{constra}
\end{equation}%
Then using equations (\ref{Gtt}), (\ref{Grr}) and (\ref{Einstein}) we find 
\begin{equation}
{\frac{1}{\alpha }}{\frac{{\partial \alpha }}{{\partial r}}}+{\frac{1}{a}}{%
\frac{{\partial a}}{{\partial r}}}={\mu ^{2}}r({X^{2}}+{A_{r}^{2}})
\label{constralpha}
\end{equation}%
To implement these equations numerically, we replace spatial derivatives
with centered differences and implement the time evolution using the
iterated Crank-Nicholson method.\cite{icn} We also put in Kreiss-Oliger
dissipation\cite{ko} for added stability. Initial data for this system is $X$
and $A_{r}$ at the initial time. Given these initial data, equations (\ref%
{constrW}), (\ref{constra}) and (\ref{constralpha}) are then integrated in
turn to obtain $W$ and the metric functions. Finally equations (\ref%
{evolveAr}) and (\ref{evolveX}) are used to produce $A_{r}$ and $X$ at the
next time step.

In addition to this unigrid code, we also perform simulations with an
adaptive mesh code. Note that our equations are quite similar to those used
by Choptuik \textit{et al} \cite{eym} to simulate critical collapse in the
Einstein-Yang-Mills system. Our adaptive code is produced by modifying the
code of reference\cite{eym} to simulate our system.

We now present the characteristic method using the coordinates of reference%
\cite{dc}. Here the metric takes the form 
\begin{equation}
d {s^2} = - {e^{2\nu}} d {u^2} - 2 {e^{\nu +\lambda}} du dr + {r^2} d {%
\Omega ^2}  \label{dcmetric}
\end{equation}
We introduce the null vectors 
\begin{eqnarray}
{l^a} = {e^{-\lambda}} {{\left ( {\frac{\partial }{{\partial r}}} \right ) }%
^a}  \label{ldef} \\
{n^a} = {e^{-\nu}} {{\left ( {\frac{\partial }{{\partial u}}}\right ) }^a} - 
{\textstyle {\frac{1 }{2}}} {e^{-\lambda}} {{\left ( {\frac{\partial }{{%
\partial r}}}\right ) }^a}  \label{ndef}
\end{eqnarray}
The matter in this coordinate system is determined by the components ${A_u},
\; {A_r}$ and ${F_{ur}}={e^{\nu +\lambda}} W$. Note that $W$ defined in this
way is the same as in equation(\ref{Wdef}) as can be seen by the fact that ${%
F^{ab}}{F_{ab}}=-2 {W^2}$ in both cases. We also introduce the quantities $%
g\equiv {e^{\nu + \lambda}}$ and ${\bar g} \equiv {e^{\nu - \lambda}}$.

For a metric of the form of equation (\ref{dcmetric}) we have 
\begin{eqnarray}
{G_{ab}}{l^a}{l^b} = {\frac{{2 {\bar g}}}{{r {g^2}}}} {\frac{{\partial g} }{{%
\partial r}}}  \label{Gll} \\
{G_{ab}}{n^a}{l^b} = {\frac{{-1}}{{g{r^2}}}} \left [ {\frac{\partial }{{%
\partial r}}} ( r {\bar g} ) - g \right ]  \label{Gnl}
\end{eqnarray}
From equation (\ref{Einstein}) it then follows that the corresponding
Einstein equations become 
\begin{eqnarray}
{\frac{{2 {\bar g}}}{{r {g^2}}}} {\frac{{\partial g} }{{\partial r}}} =2 {%
\mu ^2} {\frac{{\bar g}}{g}} {{({A_r})}^2}  \label{Ell} \\
{\frac{{-1}}{{g{r^2}}}} \left [ {\frac{\partial }{{\partial r}}} ( r {\bar g}
) - g \right ] = {W^2}  \label{Enl}
\end{eqnarray}

We now define a scalar field $\phi $ by ${\partial _{r}}\phi =\mu {A_{r}}$.
Note that this defines $\phi $ up to addition of an overall constant, since
smoothness of $\phi $ implies that ${\partial _{r}}\phi ={\partial _{u}}\phi 
$ at $r=0$. We also define the quantities $h$ and ${\bar{h}}$ by 
\begin{eqnarray}
h &\equiv &{\frac{\partial }{{\partial r}}}(r\phi )  \label{hdef} \\
{\bar{h}} &\equiv &{\frac{1}{r}}{\int_{0}^{r}}hdr  \label{hbardef}
\end{eqnarray}%
Then the solution of equations (\ref{Ell}) and (\ref{Enl}) become 
\begin{eqnarray}
g &=&\exp \left[ {\int_{0}^{r}}{\frac{1}{r}}{{(h-{\bar{h}})}^{2}}dr\right] 
\label{gsoln} \\
{\bar{g}} &=&{\frac{1}{r}}{\int_{0}^{r}}g(1-{W^{2}})dr  \label{gbarsoln}
\end{eqnarray}

Next we find an expression for the matter variable $W$ in terms of $%
h$. Contracting equation (\ref{divF}) with $l_{a}$ we find 
\begin{equation}
{\frac{\partial }{{\partial r}}}({r^{2}}W)+{\mu ^{2}}{r^{2}}{A_{r}}=0
\end{equation}%
for which the solution is 
\begin{equation}
W=-{\frac{\mu }{{r^{2}}}}{\int_{0}^{r}}r(h-{\bar{h}})dr  \label{Wsoln}
\end{equation}

We now find an evolution equation for $h$. From ${F_{ab}}=2{\partial _{[a}}{%
A_{b]}}$ it follows that 
\begin{equation}
{\partial _u}{A_r}-{\partial _r}{A_u} = g W  \label{Fur}
\end{equation}
Which from the definition of $\phi$ leads to 
\begin{equation}
{\partial _u}{\partial _r} \phi = {\partial _r} (\mu {A_u} ) + \mu g W
\label{d2u}
\end{equation}
Now from this equation, its integral with respect to $r$ and equation (\ref%
{hdef}) we find 
\begin{equation}
{\partial _u} h = {\partial _r} (r \mu {A_u} ) + \mu g r W + {\int _0 ^r}
\mu g W d r  \label{duh}
\end{equation}
Thus to find the evolution equation for $h$ we must find an expression for ${%
\partial _r} (r \mu {A_u} )$ in terms of $h$. To do this, we note that from
equation (\ref{divA}) it follows that 
\begin{equation}
{\partial _u}{A_r} + {\frac{1 }{{r^2}}} {\partial _r} \left [ {r^2} ({A_u} - 
{\bar g} {A_r})\right ] = 0  \label{divAdc}
\end{equation}
Then subtracting equation (\ref{Fur}) from equation (\ref{divAdc}) we obtain 
\begin{equation}
{\partial _r} (\mu r {A_u}) = - {\frac{1 }{2}} \mu r g W + {\frac{1 }{{2 r}}}
{\partial _r} ({r^2} {\bar g} {\partial _r} \phi )  \label{Au}
\end{equation}
Note that equation (\ref{Au}) can be integrated to yield an expression for
the remaining matter variable $A_u$ in terms of $h$. Thus, given $h$ at a
time $u$, all matter and metric variables at that time can be determined by
integrals. Now using equation (\ref{Au}) and equation (\ref{gbarsoln}) in
equation (\ref{duh}) we obtain 
\begin{equation}
D h = {\frac{1 }{{2r}}} (h-{\bar h})(g-{\bar g}) - {\frac{1 }{2}} g r {W^2}
+ {\frac{1 }{2}} \mu g r W + {\int _0 ^r} \mu g W d r  \label{evolveh}
\end{equation}
Here $D \equiv {\partial _u} - ({\bar g}/2) {\partial _r}$ is derivative
along the ingoing null direction.

The numerical method used is the same as that used in reference \cite{dgcrit}
for scalar field collapse. Initial data is given for $h$ at an initial $u$.
Equations (\ref{hbardef}), (\ref{Wsoln}), (\ref{gsoln}) and (\ref{gbarsoln})
are then integrated to find the other matter and metric variables. The
integration method is Simpson's rule for unequally spaced points, but near
the origin a Taylor series is used. Then equation (\ref{evolveh}) is used to
find $h$ at the next value of $u$. Each grid point is an ingoing light ray,
and both $h$ and $r$ are evolved along the grid points. When a grid point
reaches $r=0$, it is removed from the computational grid and when half of
the points have been lost, they are put back in between the remaining points
by using interpolation.

The scale invariance $({A_{a}},{g_{ab}}) \to (k{A_{a}},{k^2}{g_{ab}})$
of the Einstein-Maxwell system extends to the 
Einstein-Proca system.
Specifically, if (${A_{a}},{g_{ab}},\mu $%
) is a solution of equations (\ref{divF}) and (\ref{Einstein})
then ($k{A_{a}},{k^{2}}{g_{ab}},\mu /k$) is also a
solution of these equations, where $k$ is any positive constant.  This
allows us to set $\mu =1$ without loss of generality, which we do
in all runs.  Note that large $k$ can render $\mu$ negligible while
retaining the effect of the additional Proca terms.  We will encounter
this effect in our investigation of the critical behavior of this system.

\section{Results}

Runs were done on various Unix and Linux workstations and on PCs.
To test universality, we tried initial data of several forms, including 
a gaussian shape
and a sech shape.  However, here we will present only the results of runs
with the gaussian initial data.  Specifically, 
for the Cauchy codes we use the following form of initial data: $X=0$ and ${%
A_{r}}=pr\exp [-{{(r-{r_{0}})}^{2}}/{\sigma ^{2}}]$ where $p,{r_{0}}$ and $%
\sigma $ are constants.
For the characteristic code, the initial data is $\phi =p{r^{2}}\exp [-{{%
(r-{r_{0}})}^{2}}/{\sigma ^{2}}]$. In simulations, we fix $r_{0}$ and $%
\sigma $ and have $p$ as the parameter that is varied. The critical value of 
$p$ (denoted $p\ast $) is found by a binary search.

\begin{figure}
\includegraphics{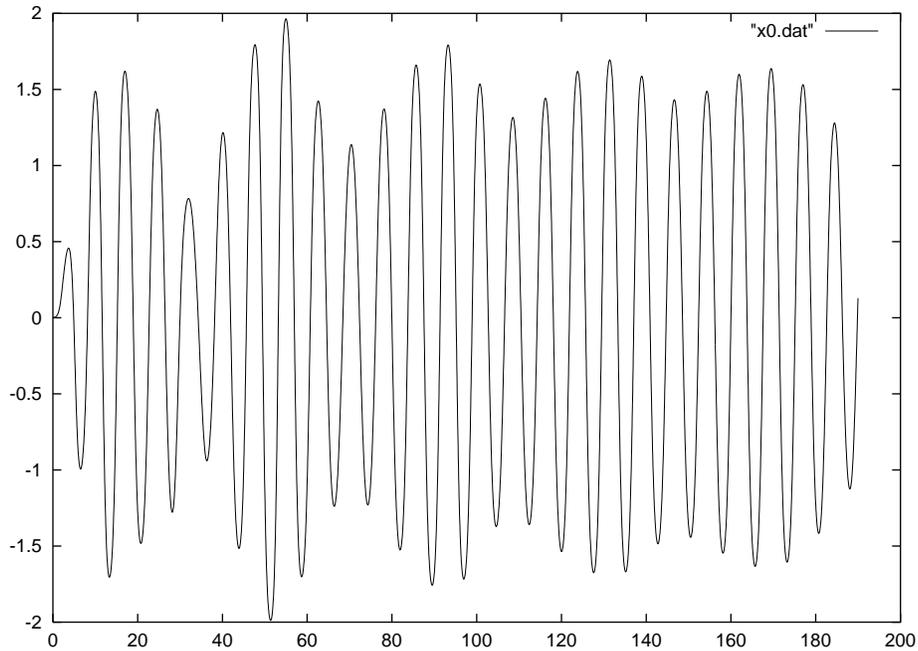}
\caption{\label{f1}$X(0)$ for the periodic critical solution}
\end{figure}

We find two different critical solutions depending 
on the value of $\sigma $%
. One is a periodic type I critical solution. Figure 1, produced using the
unigrid Cauchy code, shows $X$ at $r=0$ as a function of $t$ for this
solution. For this run, we have $\sigma =1.5, {r_0} =3.0 $ and 
$p* =0.104135195147191 $.

\begin{figure}
\includegraphics{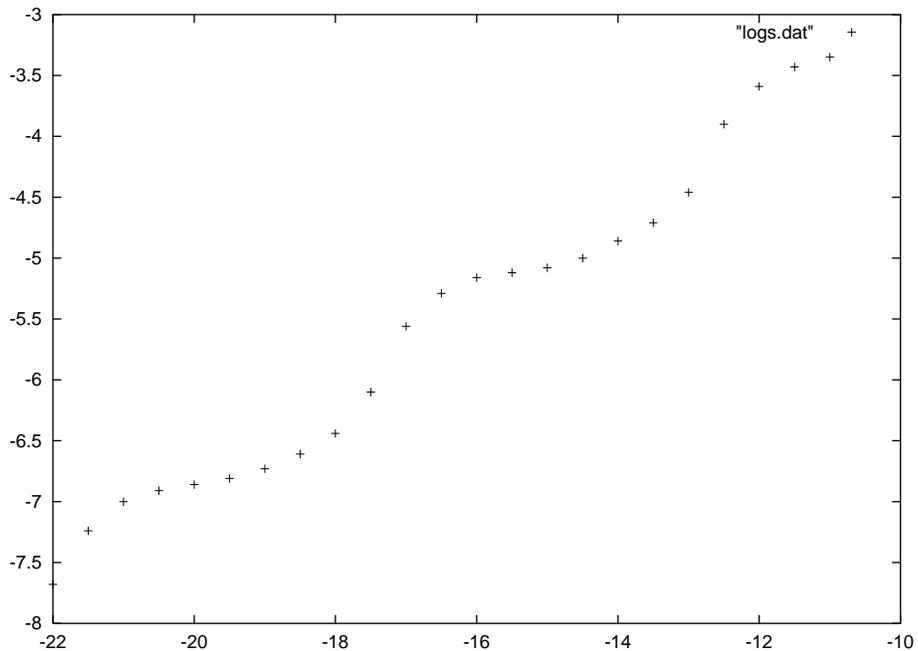}
\caption{\label{f2}$\ln M$ {\it vs} $\ln (p- p\ast )$
near the DSS critical solution}
\end{figure}

The other solution is a type II DSS critical solution which appears to be
identical to the Choptuik critical solution for a massless scalar field.
Figure 2 shows a plot of $\ln M$ vs $\ln (p-p\ast )$ for solutions above but
near the critical one. (Here $M$ is the mass of the black hole). This plot
was produced using the adaptive Cauchy code. Here, 
${r_0}=3.0, \sigma =0.5 $ and $%
p\ast = 0.134075353579$. 
For the Choptuik critical solution, the results of 
\cite{hodpiran,carsten} show that the graph of
$\ln M$ vs $\ln (p-p\ast )$
is a straight line with a periodic wiggle.  Here the slope of the line
is called $\gamma$ and the period of the wiggle is 
${T_w}=\Delta /(2\gamma )$
where $\Delta$ is the period of the DSS critical solution.  The simulations
of \cite{carsten} give $\gamma =0.374$ and $\Delta =3.4453$
which yields ${T_w}=4.61$.  We fit the data of figure (\ref{f2})
to a straight line plus a sine wave.  We find that the slope of the
line is $\gamma =0.379$ and the period of the wiggle is 
$ {T_w}=4.63$.
Thus, to the accuracy of our simulation, our DSS critical solution gives
values of $\gamma$ and $T_w$ in agreement with those of the Choptuik
critical solution.

\begin{figure}
\includegraphics{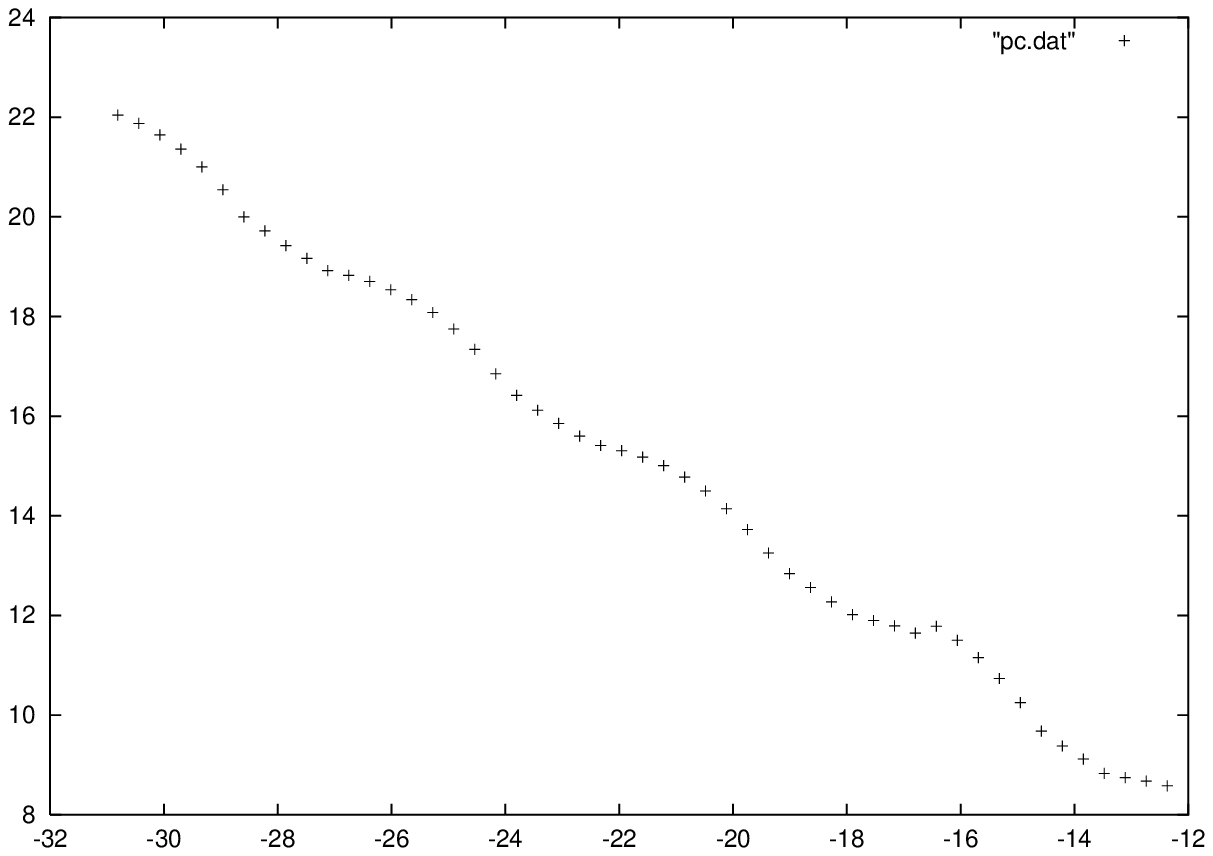}
\caption{\label{f3}$\ln {R_{\rm max}}$ {\it vs} $\ln (p\ast -p)$
near the DSS critical solution}
\end{figure}

Similarly figure 3 shows a plot of $\ln R_{\mathrm{max}}$ vs $\ln (p\ast -p)$
for solutions below but near the critical one. Here $R_{\mathrm{max}}$ is
the maximum value of the scalar curvature at the center. 
This figure was produced
using the characteristic code. Here, ${r_{0}}=2.0,\sigma =0.5$ and 
$p\ast =0.0501805022078927$.
(Note though that due to the different type of data, these parameters have
different meaning than in the Cauchy case). 
As shown in \cite{meandcomer} 
this sort of plot should also be a straight line with a periodic wiggle.
Here the slope of the line should be $-2\gamma$ and the period of the 
wiggle should be ${T_w}=\Delta /(2 \gamma)$
A fit of the data of this figure to a straight line plus a sine wave
yield that the slope of the line is $%
-2\gamma =-0.727$ which gives rise to 
$\gamma =0.363$ while the period of the wiggle is
${T_w}=4.64$. These values are again comparable to the
values of the Choptuik critical solution.

We now make a direct comparison between this critical solution and the
Choptuik critical solution for a massless scalar field. 
\begin{figure}
\includegraphics{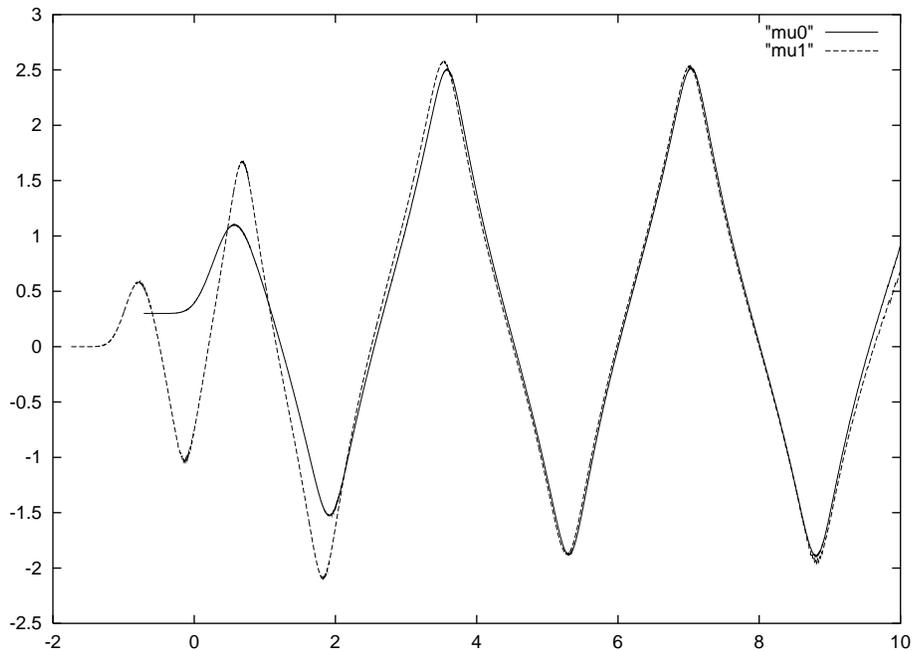}
\caption{\label{f4}$h(0)$ {\it vs} $T$ for the Proca DSS critical 
solution and the
Choptuik critical solution}
\end{figure}
The simplest way to do this is to note that our
characteristic equations (\ref{hdef} - \ref{gbarsoln}), (\ref{Wsoln}) and (%
\ref{evolveh}) formally go over to the corresponding equations for a
massless scalar field if we set the parameter $\mu $ to zero. We will return
to this point in the next section. Thus we can find the Choptuik critical
critical solution with our code by performing a binary search with $\mu =0$.
Figure 4 contains a comparison of the two critical solutions. What is
plotted is $h$ at $r=0$ as a function of $T$ where ${e^{-T}}={u\ast }-u$ and 
$u\ast $ is the value of $u$ at which the singularity forms. We use the
invariances of the two systems to choose offsets in $T$ and $h$ so that the
two solutions coincide at a maximum of $h$. Note that the two solutions
(after an initial transient has died away) are the same.

\section{Conclusions}

Given the results of reference\cite{pat} for a massive scalar field, it is
not surprising that a massive vector field has a type I periodic critical
solution. What does seem surprising is that it has a DSS critical solution
that is identical to that of a massless scalar field. For a DSS critical
solution we would expect that since length scales are becoming arbitrarily
small, that $\mu $ is becoming negligible compared to the inverse of the
relevant length scale (or rather, since we set $\mu =1$, that the relevant
inverse dimensionless length scale is becoming arbitrarily large). 

Thus the
DSS critical solution should in some sense also be a solution of the ``$\mu
\rightarrow 0$ limit'' of the equations. We have already seen how to make
sense of this limit in the case where spherical symmetry is imposed and the
system is expressed in terms of variables chosen to be similar to those of
reference\cite{dc} . We now show how to make sense of this limit more
generally using equations (\ref{divF} - \ref{Einstein}). If $A_{a}$ itself
has a smooth $\mu \rightarrow 0$ limit, then the $\mu \rightarrow 0$ limit
of equations (\ref{divF} - \ref{Einstein}) is simply the Einstein-Maxwell
equations. Instead we assume that $A_{a}$ takes the form 
\begin{equation}
{A_{a}}={\frac{1}{\mu }}{P_{a}}+\mu {Q_{a}}  \label{Asingular}
\end{equation}%
Then in order that the stress-energy have a non-singular $\mu \rightarrow 0$
limit, we must have ${\nabla _{\lbrack a}}{P_{b]}}=0$ and therefore there
must be a scalar field $\phi $ such that ${P_{a}}={\nabla _{a}}\phi $. Then
in the $\mu \rightarrow 0$ limit equations (\ref{divA}) and (\ref{Einstein})
become respectively 
\begin{eqnarray}
{\nabla _{a}}{\nabla ^{a}}\phi  &=&0  \label{wave} \\
{G_{ab}} &=&2{\nabla _{a}}\phi {\nabla _{b}}\phi -{g_{ab}}{\nabla ^{c}}\phi {%
\nabla _{c}}\phi   \label{Escalar}
\end{eqnarray}%
In other words, the $\mu \rightarrow 0$ limit of the Einstein-Proca system
becomes the Einstein-scalar system. 

Consequently it is not surprising that these two systems posess the same DSS
critical solution. We therefore see that the 
Einstein-Maxwell theory is not the $%
\mu \rightarrow 0$ limit of the Einstein-Proca system, a discontinuity
reminiscent of that observed in pure gravitation \cite{Veltvandam}. 
Indeed, since gravitation couples to all forms of energy, it couples to the
longitudinal mode of the Proca field, amplifying it during spherically
symmetric critical collapse relative to the transverse modes which become
negligible. The physics of the critical 
gravitational collapse of a Proca field
therefore becomes indistinguishable from that of a massless scalar.

We close by noting that the type I critical solution we have found is
essentially an analog for the Proca system of the soliton solution found
by Seidel and Suen\cite{edandwaimo} for the massive scalar field.  We 
therefore expect that our solution could be found directly using the 
methods of reference\cite{edandwaimo}.  
We will address this issue in a separate paper.

\section{Acknowledgements}

We thank Matt Choptuik and Frans Pretorious for helpful discussions. We also
thank Matt Choptuik for making his Einstein-Yang-Mills code available to us.
This work was partially supported by NSF grant PHY-9988790 to Oakland
University.


\begin{thebibliography}{}
\bibitem{matt} M. Choptuik, Phys. Rev. Lett. \textbf{70}, 9 (1993)

\bibitem{pat} P. Brady, C. Chambers and S. Goncalves, Phys. Rev. D \textbf{56%
}, 6057 (1997)

\bibitem{edandwaimo} E. Seidel and W. Suen, Phys. Rev. Lett. \textbf{66},
1659 (1991)

\bibitem{bob} R. Wald, \textit{General Relativity} (University of Chicago
Press, Chicago 1984)

\bibitem{iz} C. Itzykson and J. Zuber, \textit{Quantum Field Theory}
(McGraw-Hill, New York, 1980)

\bibitem{dc} D. Christodoulou, Commun. Math. Phys. \textbf{105}, 337 (1986)

\bibitem{icn} M. Choptuik, in \textit{Deterministic Chaos in General
Relativity}, edited by D. Hobill, A. Burd and A. Coley (Plenum, New York,
1994), pp. 155-175

\bibitem{ko} H. Kreiss and J. Oliger, Methods for the Approximate Solution
of Time Dependent Problems, Global Atmospheric Research Programme,
Publication Series No. 10 (1973)

\bibitem{eym} M. Choptuik, T. Chmaj and P. Bizon, Phys. Rev. Lett. \textbf{77%
}, 424 (1996)

\bibitem{dgcrit} D. Garfinkle, Phys. Rev. \textbf{D51}, 5558 (1995)

\bibitem{hodpiran} S. Hod and T. Prian, Phys. Rev. \textbf{D55}, 
440 (1997)

\bibitem{carsten} C. Gundlach, Phys. Rev. \textbf{D55}, 695 (1997)

\bibitem{meandcomer} D. Garfinkle and G.C. Duncan, Phys. Rev. \textbf{D58},
064024 (1998)

\bibitem{Veltvandam} H. van Dam and M.J.G. Veltman,  Nucl.Phys. \textbf{B22}
, 397 (1970)

\end{thebibliography}
\end{document}